\documentclass[aps,pre,twocolumn,showpacs,amsmath,amssymb]{revtex4} 
\usepackage{graphicx} 
\usepackage{dcolumn} 
\usepackage{bm} 
\usepackage[usenames,dvipsnames]{color}
\usepackage{CJK}

\begin{document}

\newcommand{\fmov}{f_\text{mov}}


\title{Multiple transient memories in sheared suspensions: \\ robustness, structure, and routes to plasticity}

\author{Nathan C. Keim\footnote{These authors contributed equally to this work}}
\email{nkeim@seas.upenn.edu}
\affiliation{Department of Mechanical Engineering and Applied Mechanics, The University of Pennsylvania}
\author{Joseph D. Paulsen$^{*}$}
\email{paulsenj@uchicago.edu}
\affiliation{The James Franck Institute and Department of Physics, The University of Chicago}
\author{Sidney R. Nagel}
\affiliation{The James Franck Institute and Department of Physics, The University of Chicago}

\date{\today}

\begin{abstract}
    Multiple transient memories, originally discovered in charge-density-wave conductors, are a remarkable and initially counterintuitive example of how a system can store information about its driving. In this class of memories, a system can learn multiple driving inputs, nearly all of which are eventually forgotten despite their continual input. If sufficient noise is present, the system regains plasticity so that it can continue to learn new memories indefinitely. Recently, Keim \& Nagel [\emph{Phys.\ Rev.\ Lett.}\ 107, 010603 (2011)] showed how multiple transient memories could be generalized to a generic driven disordered system with noise, giving as an example simulations of a simple model of a sheared non-Brownian suspension. Here, we further explore simulation models of suspensions under cyclic shear, focussing on three main themes: robustness, structure, and overdriving.  We show that multiple transient memories are a robust feature independent of many details of the model. The steady-state spatial distribution of the particles is sensitive to the driving algorithm; nonetheless, the memory formation is independent of such a change in particle correlations.  Finally, we demonstrate that overdriving provides another means for controlling memory formation and retention.  
\end{abstract}


\maketitle

\section{Introduction} 

The basic operations of memory---imprinting, reading, and erasure---can occur rapidly.  Examples are the flipping of a magnetic domain, writing a nested sequence of values in return-point memory \cite{Sethna1993}, or aging and rejuvenation in glasses \cite{Jonason1998, Zou2010}.
The situation is drastically different for pulse-duration memories in traveling charge-density waves \cite{Fleming1986, Brown1986}, where a memory can be gradually encoded into or erased from the steady-state response of the system. If a series of equal-width current pulses is applied to the system, the nonlinear voltage response eventually becomes phase-locked to the \textit{ends} of the pulses. Thus, the system has a memory of the pulse duration.

Moreover, charge-density wave carriers can simultaneously store multiple memories. If a sequence of pulses is repeatedly applied, the system will remember each of the pulse widths. Counterintuitively, all but two will be forgotten in the steady state. However, if sufficient noise is present, all of the memories can be retained \cite{Coppersmith1997, Povinelli1999}. The defining features of multiple transient memories are thus: (i) initially, the system can learn multiple inputs, (ii) under continual application of these inputs, the system will forget almost all of them, and (iii) if sufficient noise is present, the system will remember the inputs indefinitely.

Recently, Keim \& Nagel \cite{Keim2011} showed how multiple transient memories could be generalized in a generic driven disordered system with noise. Turning to classical driven systems with disorder, they noted other examples of memories stored in a steady state. For example, granular materials that are driven by tapping \cite{Knight1995} or shearing \cite{Toiya2004} assume a steady-state density that is determined by the amplitude of the driving. With prior knowledge of this compaction behavior, simply measuring the pack height after many driving cycles identifies the driving parameter uniquely---i.e.,\ the system has a single memory in its steady state.

To demonstrate that particulate systems could possess multiple transient memories, Keim \& Nagel \cite{Keim2011} employed a simplified model that was developed by Cort\'e \textit{et al.}~\cite{Corte2008} as a kinematic description of experiments of non-Brownian suspensions under cyclic, low Reynolds-number shear \cite{Pine2005, Corte2008}. In such systems, particles were found to follow irreversible trajectories before self-organizing into a reversible steady state, where the particles retrace their trajectories exactly during every subsequent shear cycle. 

In this system, a memory may be encoded by shearing a system cyclically between strains $\gamma = 0$ and $\gamma_1$. As the system self-organizes, more and more particle trajectories will become reversible for strains smaller than or equal to the strain amplitude $\gamma_1$. A memory is encoded as the sharp increase in particle irreversibility for shears infinitesimally larger than $\gamma_1$~\cite{Keim2011}. A configuration that is a reversible steady state for amplitude $\gamma_1$ is also reversible for any smaller shear amplitude, $\gamma<\gamma_1$. Hence, if multiple memories are stored in the system by shearing to multiple different amplitudes $\{\gamma_1, \gamma_2, ... \gamma_n\}$, $\gamma_n$ being the largest, then once the system reaches the final reversible steady state (i.e.,\ complete reversibility for shears up to $\gamma_n$), the smaller memories will have been forgotten. There then remains only the onset of irreversibility at $\gamma_n$.  We note here that the system has effectively two memories in the final steady-state configuration (in analogy with the charge-density-wave system).  One of these is, of course, at $\gamma_n$; the other is the maximum strain in the opposite direction, which in this work is fixed at $\gamma = 0$.  These values mark the two endpoints of the driving during the training cycles.

Crucially, if noise is present in the system, the suspension can never reach its final fixed point where all the particle trajectories would be fully reversible up to a strain amplitude $\gamma_n$. Thus the system explores a space of metastable states, and retains signatures of the inputs indefinitely. Because it is effectively maintained in a permanent transient, if the pattern of inputs is changed, the system will gradually acquire new memories and shed the old ones.  The ability to retain a memory of the small inputs is never lost.  This is a concrete example of the emergence of plasticity in memory formation.

While multiple transient memories are likely a generic phenomenon, a rigorous understanding of the conditions under which they are feasible has yet to have been expressed. Here, we address some of the issues underlying this question, by further exploring the sheared suspension system and focussing on three main themes: robustness, the structure of memories, and overdriving. We begin by describing the basic simulation algorithm in section II. In section III, we show that multiple transient memories are robust and are present under several different training algorithms. We then show that forgetting is gradual---a crucial property that distinguishes transient memories from other classes of memory. We demonstrate this over a range of parameters and algorithm variants by reading out a small memory after a single large shear has been applied. In section IV, we examine the spatial structure of particle configurations that possess memories in the Cort\'e algorithm, and we show how this and other effects lead to a broadening of the memory signature. 

Finally, in section V, we examine sheared suspensions driven above the critical amplitude for irreversibility, $\gamma_c$, the largest strain amplitude at which the system can self-organize to a reversible steady state~\cite{Corte2008}. We show that even in the case of overdriving, memories can be retained in the system just as for smaller amplitude inputs.  Thus memories are not necessarily relegated to small amplitudes.  Morevover, driving above $\gamma_c$ can be harnessed as a source of noise that allows memories to be retained indefinitely. 

The picture that emerges from this work is that multiple transient memories are a robust feature manifest in a range of simple models of suspensions under cyclic, low Reynolds-number shear, that the details of memory retention and erasure can be understood from the spatial structure of the particles, and that overdriving can provide another avenue for controlling memory formation.


\section{Description of Simulations} 
\label{sec:algorithm}

We use a variety of models to study the rearrangements of particles in a viscous, non-Brownian suspensions under cyclic low Reynolds-number shear. One of these models was designed by Cort\'e \textit{et al.}~\cite{Corte2008}. A slight variant was used by Keim \& Nagel \cite{Keim2011} to show multiple transient memories in sheared suspensions. Although the models are very simplified, they are justified on the basis that they reproduce many behaviors seen in previous experiments \cite{Corte2008}.

\begin{figure}
\includegraphics[width=2.5in]{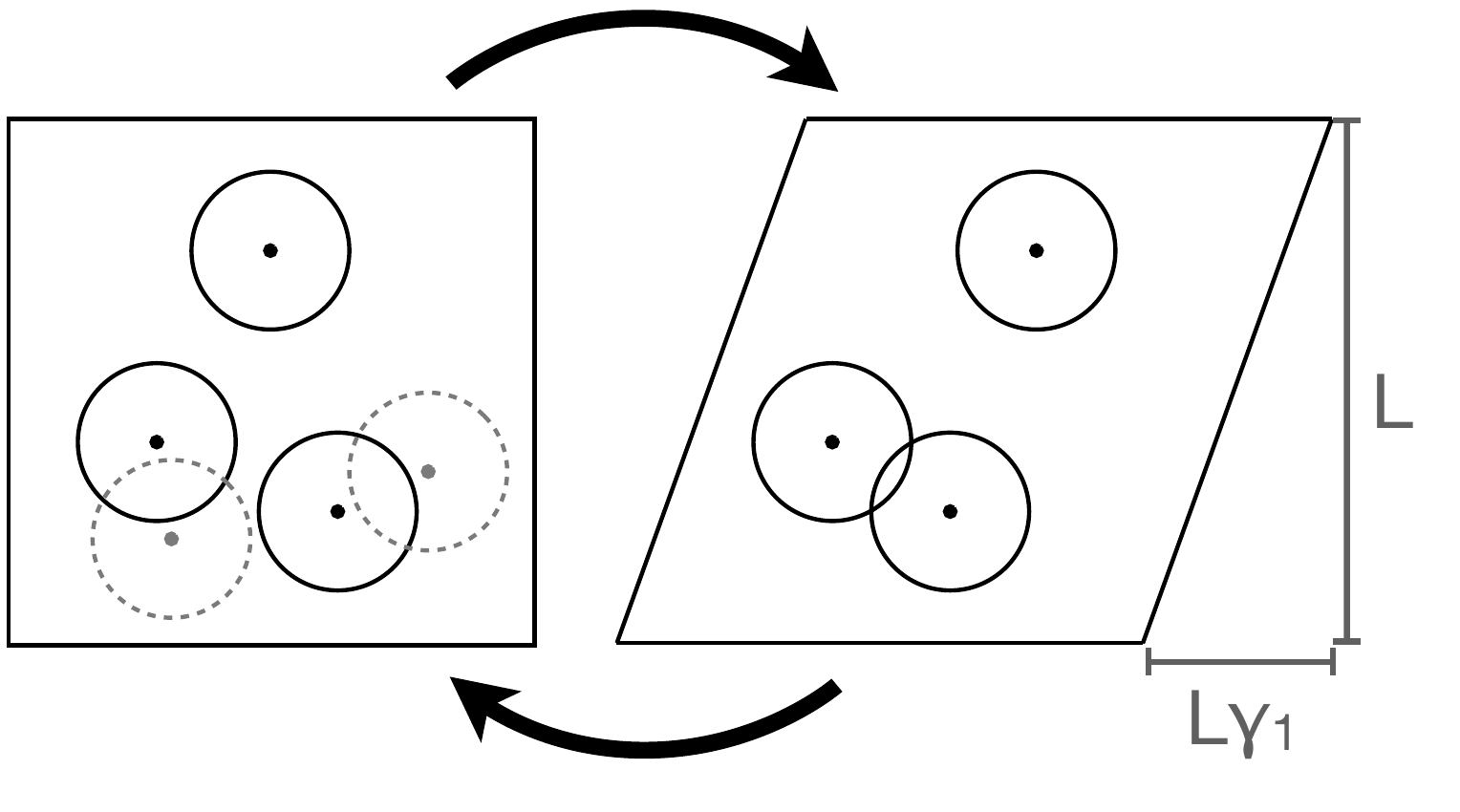}
\caption{
Simulation algorithm of Cort\'e \textit{et al.}~\cite{Corte2008}. $N$ discs of diameter $d$ are randomly placed in a square box with area fraction $\phi$ (left). During each cycle, the box is sheared affinely by translating the particle centers a distance $\Delta x= \gamma_1 y$ (right). They are then returned to their previous positions (left). Any particle pairs that overlap at any point during the cycle are given small displacements (left, dashed circles), which determine the new positions for the next shear cycle.  Figure adapted from \cite{Corte2008}.
\label{algorithm}
}
\end{figure}

The algorithm of Cort\'e \textit{et al.}~\cite{Corte2008} consists of three steps, illustrated in Fig.\ \ref{algorithm}: (1) $N$ discs of diameter $d$ are randomly placed in a square box of area $A_{box}$ with area fraction $\phi=N\pi (d/2)^2/A_\text{box}$. We choose $\phi=0.2$ and use periodic boundary conditions at the box edges. (2) The box is sheared affinely in the $x$ direction to an amplitude $\gamma_1$ so that a particle is translated a distance $\Delta x = \gamma_1 y$, where $y$ is the vertical coordinate of the particle center.  The particles are then returned to their original positions. (3) Any discs that overlap each other at any point during the shear (i.e.,\ if their centers come within a distance $d$) are given small, random displacements (or ``kicks''), once they have been returned to their unsheared positions. In each variant of the algorithm that we study, a different choice for the displacements is prescribed.

This very simple model reproduces several important features of the sheared-suspension experiment. First, for small shear amplitude (where the definition of ``small'' depends on the area fraction), the system will eventually fall into a reversible steady state (i.e.,\ no particle pairs come within a distance $d$ of each other during the shear to $\gamma_1$). Second, for large enough shears, the particles never find a reversible configuration (on the timescale of the simulation), and instead each particle undergoes a random walk indefinitely. Third, there is a critical $\gamma_c$ (which depends on the area fraction, $\phi$) that separates these two regimes.  These features were also seen in the experiment \cite{Corte2008}.

Recently, Keim \& Nagel \cite{Keim2011} showed how such a system that is driven below its critical amplitude ($\gamma<\gamma_c$) can store multiple memories in its transient state, or in a steady state when random noise is added. In the current study, we will consider the different choices for the kicks applied to the overlapping particles, as well as different forms of driving.  We also consider the effect of overdriving, that is the effect of shears of amplitude $\gamma>\gamma_c$. In the present work $\phi=0.2$, and empirically $\gamma_c = 4.0$. Without loss of generality, we set the particle diameter $d=1$.


\section{Robustness of Multiple Transient Memories} 

The capacity of a system with many metastable states and many degrees of freedom to ``learn'' multiple memories under cyclic driving appears to require dynamics with very few essential features~\cite{Keim2011}. Here we test and study the generality of this learning behavior under different algorithms. First, we evaluate several variants of the original simulation algorithm. We then consider one aspect of the kinematics that is necessary for multiple transient memories: the persistence of a memory over many cycles.

\subsection{Dependence on Kinematics} 
\label{theme1}

\begin{figure}
\includegraphics[width=2.9in]{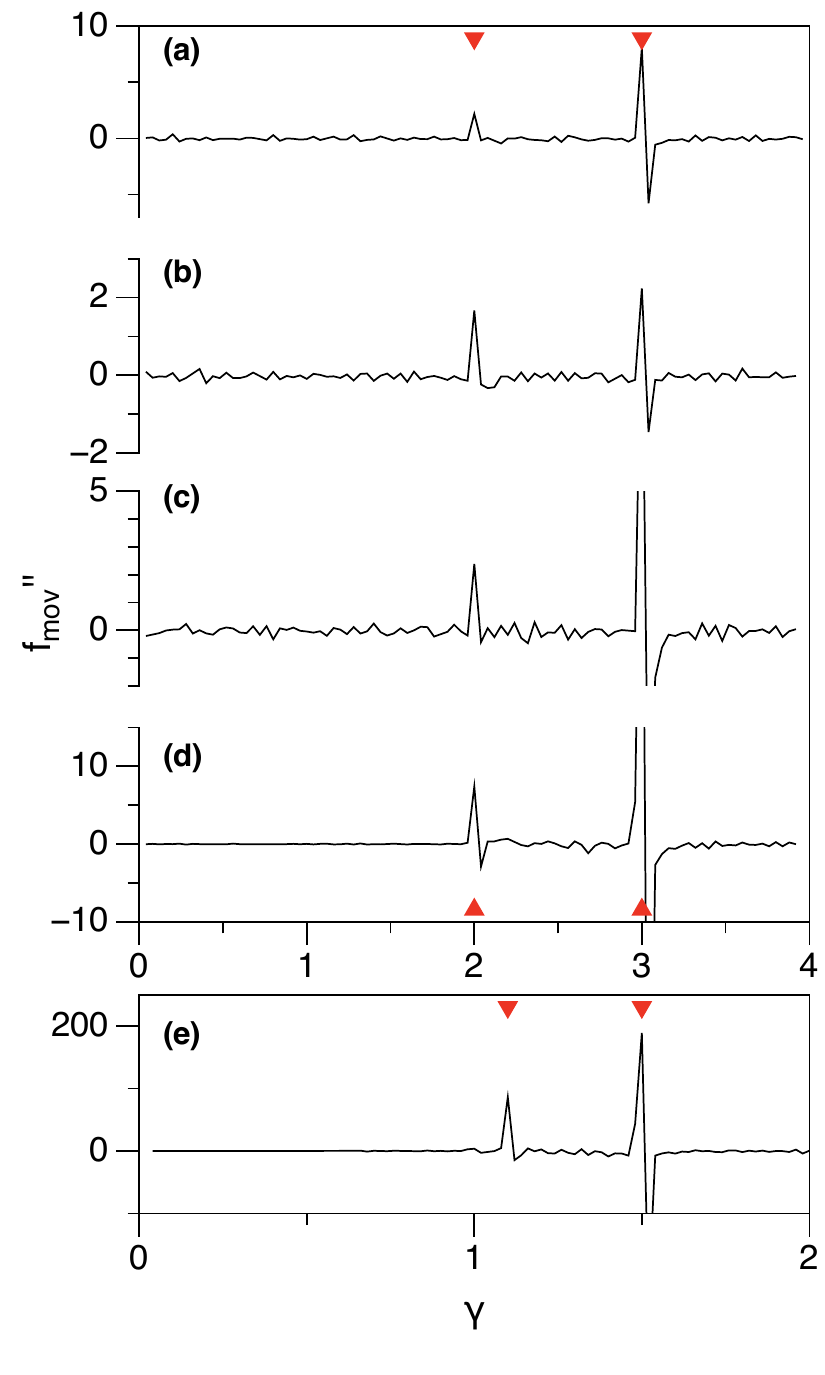}
\caption{
    (color online) Multiple memories observed using variants of the kinematics. Memory signal, $f_\mathrm{mov}''$, is plotted versus trial strain, $\gamma$. Red triangles above and below curves indicate values used in the training pattern. \textbf{(a)} Variant~\ref{variant:corte} in text, with an isotropic random kick for each collision, after $\sim 10^4$ cycles (average of 12 runs)~\cite{Corte2008}. \textbf{(b)} Variant~\ref{variant:tko}: each particle kicked at most once per cycle, regardless of number of collisions~\cite{Keim2011}, after $\sim 10^4$ cycles (12 runs); \textbf{(c)} Variant~\ref{variant:momcons}: equal and opposite kicks, after $\sim 10^4$ cycles (12 runs, actual height of clipped peak, $13$, trough, $-8$); \textbf{(d)} Variant~\ref{variant:repulsion}: equal repulsive kicks after $\sim 10^3$ cycles (12 runs, actual height of clipped peak, $62$, trough, $-62$); \textbf{(e)} Variant~\ref{variant:dilation}: Isotropic repulsion (simulating particle swelling) after 200 cycles (5 runs, height of trough: $-234$). A single cycle is plotted for each variant, taken from a period in that system's evolution when both memories are observable after each cycle, independent of which $\gamma_i$ was just applied.
\label{fig:variants}
}
\end{figure}

To test whether memories are present after some number of cycles, we evaluate $\fmov''(\gamma) \equiv d^2 \fmov / d\gamma^2$, where $\fmov(\gamma)$ is the fraction of particles that collide in the course of a trial deformation by $\gamma$. The quantity $\fmov(\gamma)$ is cumulative, including all particles colliding at shears up to $\gamma$, whereas $\fmov'(\gamma)$ measures the set of particles colliding at $\gamma$ specifically. The ``training'' from repeated shearing therefore depletes $\fmov'$ just below the training value and enhances it above. This change is observed as a peak in $\fmov''$ for each $\gamma_i$ when multiple transient memories are present, as in Fig.~\ref{fig:variants}(a). To find $\fmov''$, we sample $\fmov$ at intervals of 0.04 in $\gamma$ (except in Variant~\ref{variant:dilation} below). We then take the discrete first and second derivatives by subtracting the values at neighboring points. 

Our simulations suggest that the multiple-memory behavior is insensitive to qualitative changes in the kinematics. We have assessed the robustness of the multiple-memory behavior under 5 variants of the kinematics, all with $\phi = 0.2$, $N=10^4$. We average over multiple runs, each with a different random initial condition. Representative results from each variant are plotted in Fig.~\ref{fig:variants}.

\begin{enumerate}
    \item Original kinematics of Cort\'e \textit{et al.}~\cite{Corte2008} (see Sec.~\ref{sec:algorithm}): \label{variant:corte} If, during a shear deformation cycle, two particles overlap, then each particle is given a kick with random direction, and magnitude drawn from a uniform distribution on [0, $\epsilon$] where here $\epsilon = 0.005$. 
Particles are sheared cyclically with $\gamma_1 = 2$ and $\gamma_2 = 3$, and the smaller shear is applied 5 times for each application of the large shear.  Thus the pattern of shears is $3, 2, 2, 2, 2, 2$, and in Fig.~\ref{fig:variants}(a) this 6-cycle pattern is repeated for a total of $\sim 10^4$ cycles.

    \item ``Tag-kick-once''~\cite{Keim2011}: \label{variant:tko} We use the same algorithm as in  Variant~\ref{variant:corte} except that each colliding particle is kicked only once, independent of the number of other particles that it overlapped. The kick is drawn from the same distribution as in Variant~\ref{variant:corte}. 

    \item Momentum conservation: \label{variant:momcons} Kicks to interacting particles are still drawn from the same isotropic random distribution as Variant~\ref{variant:corte}, and all parameters are the same. However, each particle in a colliding pair receives an equal and opposite kick. As in Variant~\ref{variant:corte}, kicks from multiple collisions are applied additively. Thus if any subset of the system does not interact with particles outside that subset, its center of mass will not move.

    \item Repulsion: \label{variant:repulsion} Particles in an interacting pair receive kicks away from each other, as defined by their positions at $\gamma = 0$. Kick magnitudes are equal, and are taken from the same distribution as in Variant~\ref{variant:corte}. Notably, we find that the system evolves at an approximately normal pace (including forgetting of the memory of the smaller shear value), yet it usually does not reach a reversible state within $2 \times 10^6$ cycles (for these parameters). 

    \item Dilation and repulsion: \label{variant:dilation} Instead of a shearing motion, kinematics simulate uniform cyclic swelling of the particles to a diameter $\gamma d$.  As in the variants described above, we used a training pattern with two values of $\gamma$, with the smaller value repeated 5 times for each application of the larger value: $\gamma_i =$ 1.5, 1.1, 1.1, 1.1, 1.1, 1.1.  We use $\epsilon = 0.01$. Kicks to overlapping particles are repulsive and have equal randomly-selected magnitude, as in Variant~\ref{variant:repulsion}. For this algorithm we find that $\gamma_c \approx 1.8$, and we sample $\fmov$ at intervals of 0.02.

\end{enumerate}

As seen in Fig.~\ref{fig:variants}, each of these variants supports multiple transient memories. Our results thus show that the phenomenon is not sensitive to how particles interact, or even to the geometry of deformation.


\subsection{Gradual Forgetting} 
\label{sec:gradual}

One important feature of transient memories is that the process of forgetting is gradual. This is observed in simulations of non-Brownian sheared suspensions \cite{Keim2011} and simulations of traveling charge-density waves \cite{Coppersmith1997, Povinelli1999}. For sheared suspensions, gradual forgetting is the property that a memory of $\gamma_i$ will persist even after one or more applications of a larger shear, $\gamma_j>\gamma_i$. This feature helps distinguish transient memories from other classes of memories. For example, in return-point memory in ferromagnets, smaller memories are wiped out the instant a larger field is applied \cite{Barker1983, Sethna1993}. The same holds for aging and rejuvenation in glasses; a memory of a temperature $T_i$ is erased as soon as the glass is heated above $T_i$ \cite{Jonason1998, Zou2010}.

In ref.~\cite{Keim2011}, the kick size was small ($0.005 \leq \epsilon \leq 0.1$). Here we show that forgetting is still gradual (albeit more rapid) when larger kicks are used ($0.01\leq \epsilon \leq 1$). Using the Cort\'e algorithm (Variant~\ref{variant:corte}), we train an initially randomized system ($N=10^5$, $\phi=0.2$) with a single strain amplitude, $\gamma_1=2$, to a reversible steady state. Then, we apply a single shear of strain amplitude $\gamma_2=3$, and attempt to read out the memory at $\gamma_1$. We average multiple systems together to improve the signal-to-noise ratio (between 3 and 400 systems), and we sample $\fmov$ at intervals of $\Delta \gamma = 0.1$.

\begin{figure}[bt] 
\centering 
\begin{center} 
\includegraphics[width=2.5in]{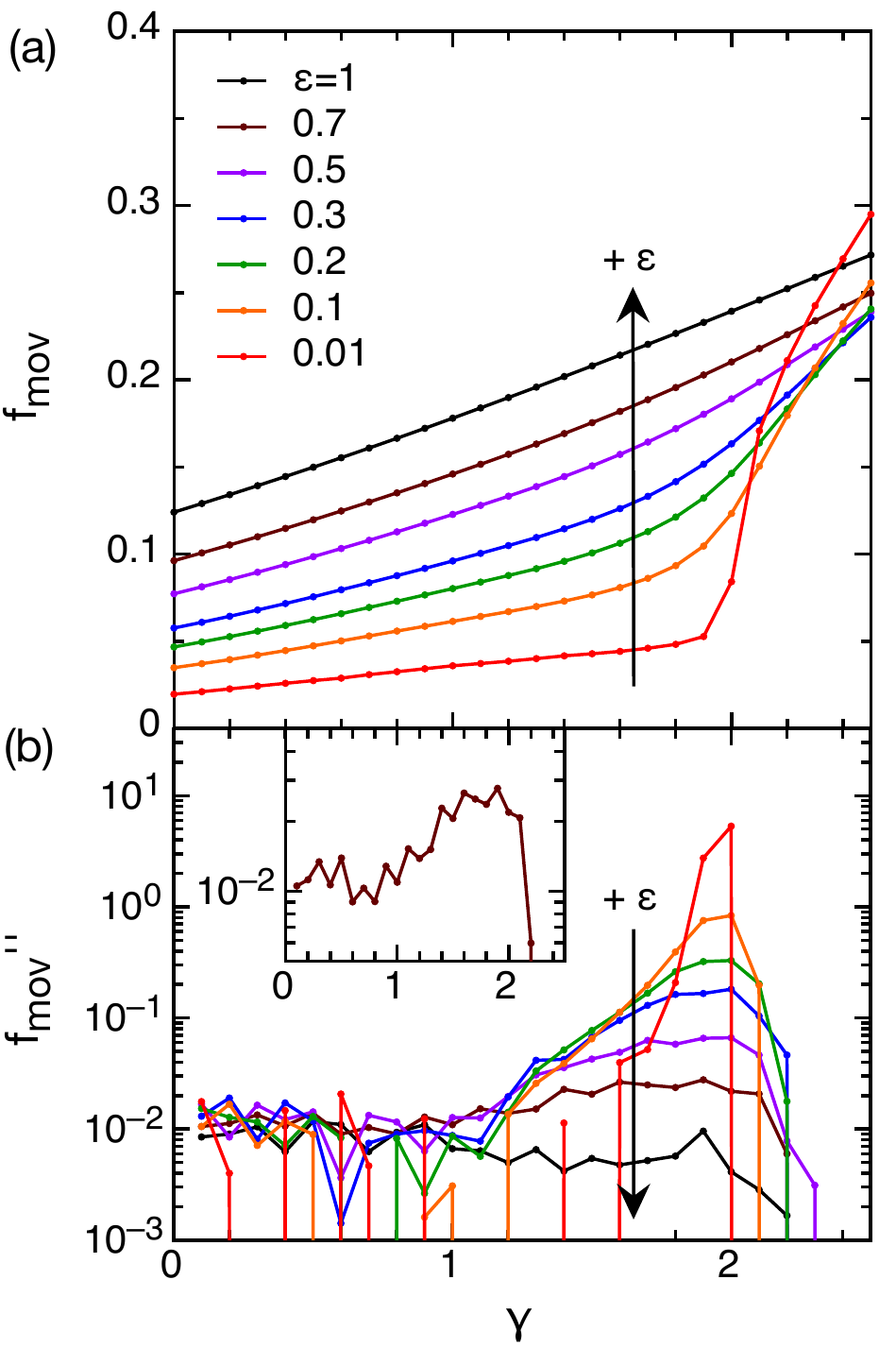} 
\end{center}
\caption{(color online) Gradual forgetting of transient memories. \textbf{(a)} $\fmov$ versus $\gamma$ for systems that were driven once at a shear amplitude $\gamma_2=3$ after having been completely trained at $\gamma_1=2$. The kick size, $\epsilon$, is given in the legend, and the arrows show the direction of increasing $\epsilon$.
\textbf{(b)} The curvature of the data, $\fmov''$, shows that the memory at $\gamma_1=2$ is still present, even for large $\epsilon$. As $\epsilon$ is increased, the peak becomes broader and shorter. \textit{Inset}: The data for $\epsilon=0.7$ show that the memory can be read out even at this large a kick size. All data were taken from averaging over multiple initial configurations.}
\label{gradualforgetting}
\end{figure}

We show the results in Fig.\ \ref{gradualforgetting}. We plot $\fmov$ and its curvature, $\fmov''$, versus strain for each value of $\epsilon$ tested. A peak is clearly visible for small kicks. As $\epsilon$ is increased, the peak becomes broader and shorter. Nonetheless, there is still an identifiable peak in $\fmov''$ for kicks as large as $\epsilon=0.7$, as shown in the inset to Fig.\ \ref{gradualforgetting}(b). (Examining the $\fmov$ data at $\epsilon=0.7$ by eye, one might not identify the memory at $\gamma_1=2$. It is remarkable that the memory can be easily identified by taking two derivatives.)
The intuitively appealing picture that emerges is that the smaller the kick size, $\epsilon$, the slower the erasing of the memory at the lower shear value.

One way in which the memory might be surviving is due to the distribution of random kicks used. In the Cort\'e algorithm, each colliding particle is given a kick with a random size chosen uniformly between $0$ and $\epsilon$. Thus, even if $\epsilon$ is large, some particles are given small kicks, and these particles alone might store the smaller memory. To investigate this, we ran simulations where a kicks of fixed size $\epsilon$ are applied (in random directions) to colliding particles. In this case, we found that the memory at $\gamma_1=2$ was still present up to $\epsilon=0.3$.

We tested three other variants of the Cort\'e algorithm, to further test the generality of gradual forgetting. In each case the system is first trained to a reversible state at $\gamma_1=2$ and then a single shear of $\gamma_2=3$ is applied. The algorithms below were used in both the initial training to the reversible state, and in the final, disrupting shear. 

\begin{enumerate}
    \setcounter{enumi}{5}
    \item Annealing at zero shear: After every cycle, overlapping particles are given random kicks until the system reaches a state with no overlaps (in the unsheared configuration). Gradual forgetting was observed at $\epsilon=0.2$ (no other values were tested). 

    \item Pure shear: Particle centers are sheared along trajectories given by: $\Delta x=x(\cosh(\frac{1}{2}\gamma)-1)+y\sinh(\frac{1}{2}\gamma)$, $\Delta y=y(\cosh(\frac{1}{2}\gamma)-1)+x\sinh(\frac{1}{2}\gamma)$. This is in contrast to simple shear, where $\Delta x=\gamma y$ and $\Delta y=0$, as in the Cort\'e algorithm. (The relationship between pure shear and simple shear is more easily understood in terms of the instantaneous flow field, which is defined in terms of the shear rate, $\dot{\gamma} \equiv d\gamma(t)/dt$. Pure shear has velocity components $v_x=\frac{1}{2}\dot{\gamma} y$, $v_y=\frac{1}{2}\dot{\gamma} x$, whereas simple shear has $v_x=\dot{\gamma} y$, $v_y=0$.) Here gradual forgetting was observed at $\epsilon=0.3$ (larger $\epsilon$ were not tested). 

    \item Dilation: The particles are swelled to radius $\sqrt{1+\gamma/\pi}$ (this radius is chosen so that the area of the interaction region matches that for shearing). Gradual forgetting was observed for $\epsilon=0.1$.
    
\end{enumerate}

Thus, the property of gradual forgetting is not relegated to small kicks and does not depend on the specific properties of the Cort\'e algorithm. It is a general feature of a range of kinematics of non-Brownian particles.



\section{Structure of a memory} 
\label{theme2}

\subsection{Pair correlation function}

\begin{figure}[bt] 
\centering 
\begin{center} 
\includegraphics[width=3.4in]{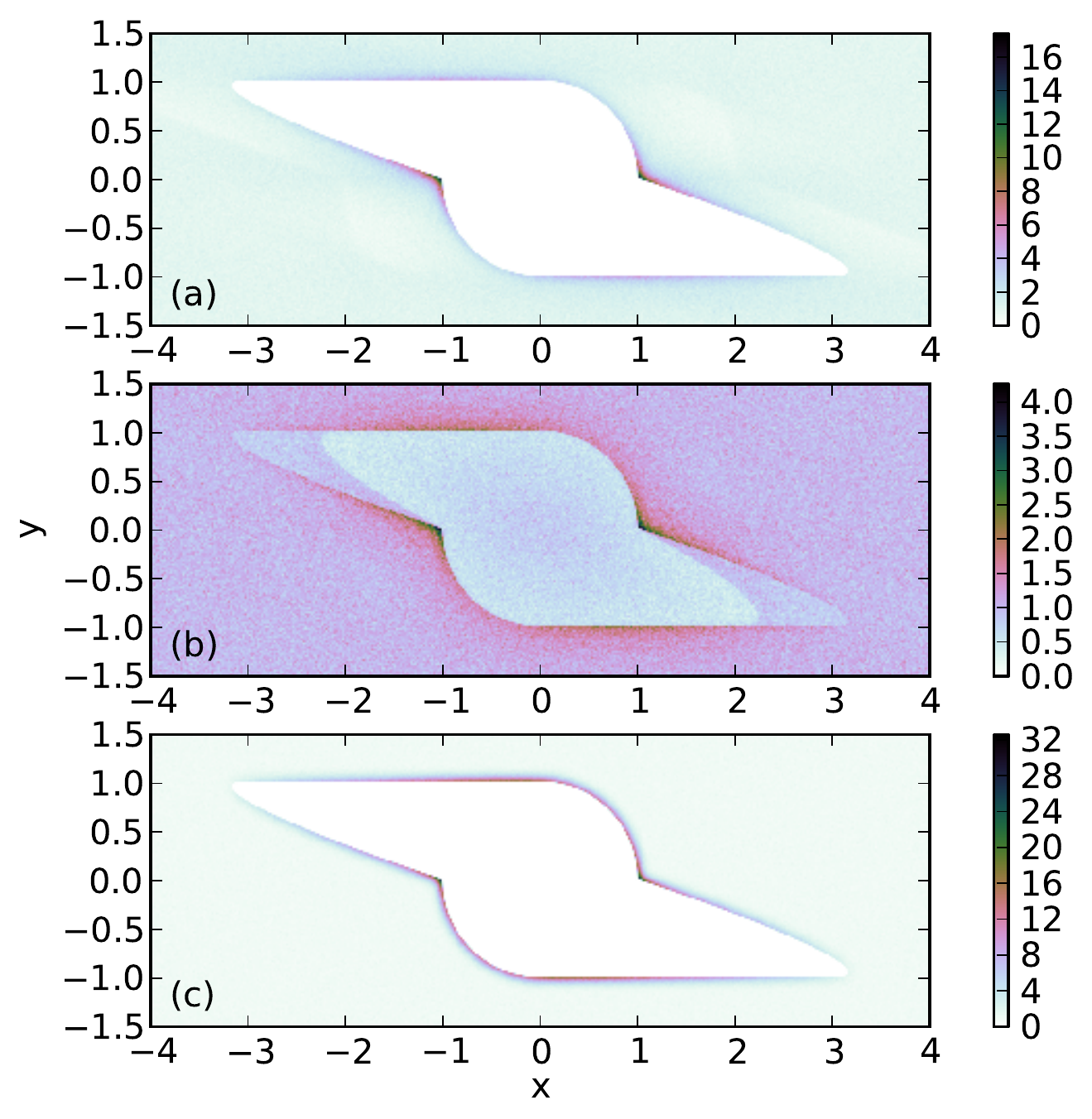} 
\end{center}
\caption{(color online) Pair correlation function, $g(x,y)$, of trained systems, where $x$ and $y$ are the flow and gradient axes, respectively. \textbf{(a)} Complete memory of $\gamma_1=3$. All particles lie outside the excluded region created by the simple shear deformation. $g(x,y)$ has peaks at $(\pm 1,0)$ where there are cusps in the shape of the excluded region . \textbf{(b)} Two memories at $\gamma_1=2$ and $\gamma_2=3$, trained with 102 cycles following the pattern: $3,2,2,2,2,2$. Particles accumulate on the boundaries of the two regions given by eqn.\ \ref{excludedregion} for $\gamma_1=2$ and for $\gamma_2=3$. This accumulation marks the formation of two memories.  Data in (a, b) were averaged over 200 systems with $N=10^4$, $\phi=0.20$, and $\epsilon=0.1$. \textbf{(c)} Isolated pairs of particles trained with $\gamma_1=3$ (averaged over $1.22\times 10^7$ pairs, with $\epsilon=0.1$). The pair structure for the full simulation (a) and for isolated pairs (c) are qualitatively similar. Differences are caused by three-body effects.}
\label{structure}
\end{figure}

The two-dimensional (2D) pair correlation function, $g(x,y)$, affords us one method of characterizing the structure of a system with one or more memories. The pair correlation function is proportional to the probablity that two particles will be separated by the displacement $(x,y)$ and is normalized so that $g(x,y)=1$ for a uniform distribution of particles.  A complete memory of an amplitude $\gamma_1$ imposes a constraint on $g(x,y)$. Because there can be no overlapping particles in the unsheared configuration, there can be no particle centers within a radial distance 1 from the origin. Furthermore, a complete memory entails that no particle centers fall in this region when the system is sheared continuously up to $\gamma_1$. This constraint creates a propeller-shaped region where $g(x,y)=0$ for a complete memory, as shown in Fig.\ \ref{structure}(a). The data are averaged over 200 systems with $N=10^4$, $\phi=0.20$, and $\epsilon=0.1$. This excluded region is the set of points satisfying:
\begin{equation}
-\sqrt{1-y^2} \mp \gamma y \leq \pm x \leq \sqrt{1-y^2},
\label{excludedregion}
\end{equation}

\noindent  for $0 \leq \pm y \leq 1$.

Two excluded regions for different values of $\gamma_i$ will nest: if $\gamma_1<\gamma_2$, then the excluded region for $\gamma_1$ is a subset of the excluded region for $\gamma_2$. This is equivalent to an observation made in the introduction and in ref.~\cite{Keim2011}: in the algorithms explored here, complete reversibility at one value of $\gamma$ dictates complete reversibility for any smaller strain amplitude. When the system has two or more memories, $g(x,y)$ is depleted of particles to a varying degree in each of the corresponding excluded regions. We show this in Fig.\ \ref{structure}(b) for a system driven at two amplitudes: $\gamma_1=2$ and $\gamma_2=3$. The data are averaged over 200 systems with $N=10^4$, $\phi=0.20$, and $\epsilon=0.1$.

In both the partial and complete memories, the pair correlation function has peaks at $(\pm 1,0)$. The peaks can be understood by the following argument. If two particles collide during a shear cycle, they will contribute to the pair correlation function at two points \textit{inside} the excluded region. These particles undergo a random walk until they exit the excluded region. Thus, isolated pairs of particles undergo a 2D diffusion process with an absorbing boundary. The curvature of the boundary of the excluded region of $g(x,y)$ influences the local density of particles exiting the region at that point---large positive curvatures mean few particles will escape there, while large negative curvatures lead to high accumulation at the boundary. The boundary of the excluded region of $g(x,y)$ has concave cusps at $(\pm 1,0)$ (i.e.,\ the curvature is negative infinity), which leads to a divergent density of particles at these points in the limit $\epsilon \to 0$.

This picture ignores three-body effects, which could spread out these peaks. To test the relative strength of three-body effects, we mimic the extreme low area-fraction limit of the Cort\'e kinematics. Namely, we simulate an ensemble of isolated pairs of particles. We evolve the pairs by giving both particles a random kick if they collide (which is equivalent to leaving one particle at the origin and giving the other particle \emph{two} random kicks). The systems are each evolved to a reversible state. We show the results in Fig.\ \ref{structure}(c), where we plot $g(x,y)$ accumulated over $1.22\times 10^7$ systems of 2 particles with $\epsilon=0.1$. 
The diffusion picture qualitatively captures the peaks in the pair correlation function at $(\pm 1,0)$. Comparing $g(x,y)$ from the diffusion simulation to the full simulation indicates that three-body effects subdue the concentration of particles along much of the boundary of the excluded region, except near $(\pm 1,0)$.

\subsection{Memory Step Function}

\begin{figure*}[bt] 
\centering 
\begin{center} 
\includegraphics[width=7.0in]{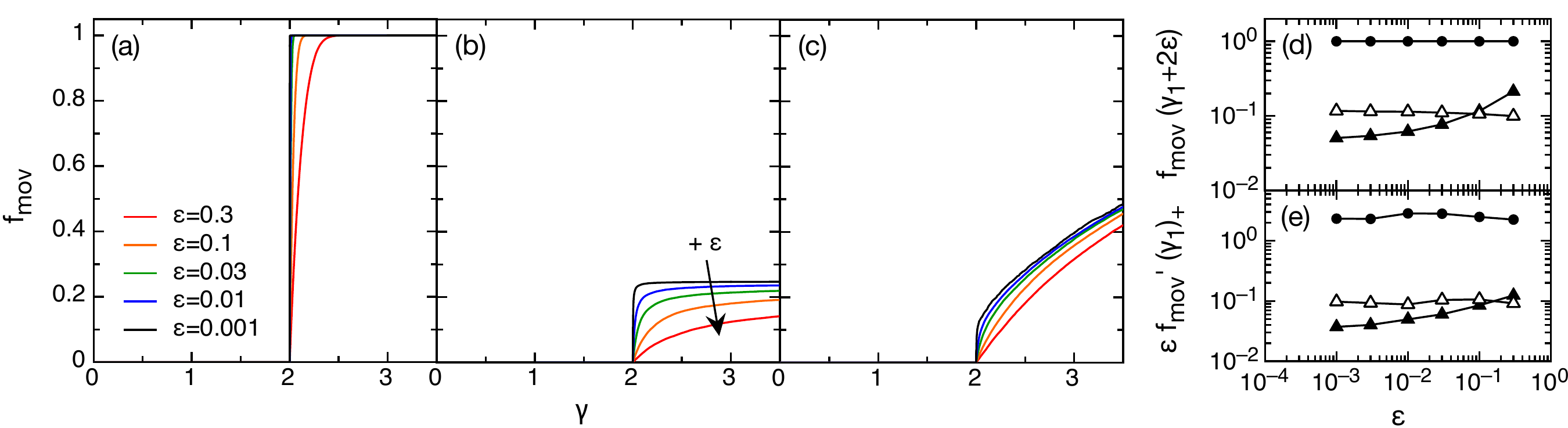} 
\end{center}
\caption{(color online)  Step function in the memory. \textbf{(a,b,c)} $\fmov$ versus strain, $\gamma$, for systems with complete memory of $\gamma_1=2$, versus kick size, $\epsilon$, trained with different algorithm variants: \textbf{(a)} radial swelling of isolated pairs of particles to diameter $\gamma_1$, \textbf{(b)} shearing of isolated pairs of particles, and \textbf{(c)} Cort\'e algorithm. In (a), as $\epsilon$ approaches zero, the memory approaches a step function in $\fmov$, since all pairs are within $2\epsilon$ of colliding at the end of the simulation. In (b), the step function is smoothed out and the height of the jump shortened. In (c), three-body effects shorten and smooth the jump even more. In all cases, a smaller kick size, $\epsilon$, steepens $\fmov$ at $\gamma_1$. Data in (a,b) are from $5\times 10^4$ isolated pairs of particles. Data in (c) are single systems of $N=10^5$ particles (except for $\epsilon=0.001$, where $N=10^4$). The arrow in (b) shows the direction of increasing $\epsilon$, which is the same in (a) and (c). \textbf{(d)} The size of the jump in $\fmov$ at $\gamma_1=2$ for radial swelling of isolated particles ($\bullet$), for shearing of isolated particles ($\vartriangle$), and for the Cort\'e algorithm ($\blacktriangle$). \textbf{(e)} Slope of $\fmov$ on the right-hand side of $\gamma_1$, scaled by $\epsilon$. In both (d) and (e), the data appear to plateau at a finite value for all three algorithms in the limit of small $\epsilon$. The plateau value indicates the degree to which the behavior is approximated by a step function.}
\label{memorystep}
\end{figure*}

When a system stores a partial or complete memory of some amplitude, many pairs of particles are nearly touching at that shear value. That is, the training has pushed the particles just far enough so that they do not collide at the shear amplitude used in the training.  In examining the pair correlation function, $g(x,y)$, we see that this observation can be stated in the following way: there is a high concentration of particles within a distance $2\epsilon$ from the boundary of the excluded region in $g(x,y)$. Therefore, many particle pairs that behave reversibly for a shear to $\gamma_1$, will collide when the system is driven to $\gamma_1+2\epsilon$. In the case of a complete memory, this means that $\fmov(\gamma_1)=0$, whereas $\fmov(\gamma_1+2 \epsilon)$ is finite. Consequently, one might expect a step function in $\fmov$ at $\gamma_1$ for small $\epsilon$. Counter to this expectation, none of the data reported by Keim \& Nagel \cite{Keim2011} show a step function in $\fmov$ at the memory amplitudes.

To understand the apparent absence of a step function in $\fmov$, we consider a simpler system, where a step-function in $\fmov$ is clearly present. In Fig.\ \ref{memorystep}(a), we show $\fmov$ for simulations of $5\times 10^4$ isolated pairs of particles that initially overlap, trained by cyclic swelling (to diameter $\gamma$). We vary the kick size from $\epsilon=0.3$ down to $0.001$. The $\fmov$ data approach a step-function at $\gamma_1=2$ in the limit of small $\epsilon$.

The step function is widened for isolated pairs of particles under \emph{shear}: if a particle lands near the points $(\pm 1, 0)$, a very large shear may be necessary for a collision. For example, a particle that is displaced to the point $(1+\epsilon, 0)$ will not collide with its partner, even for arbitrarily large strain amplitudes. If displaced to $(1+\epsilon,\delta)$ (where $\delta>0$ is small), a shear of $\gamma \approx \epsilon/\delta$ is required for a collision---the smaller the displacement off the y-axis, $\delta$, the larger $\gamma$ must be for a collision. Although this effect is specific to a vanishingly small region of configuration space of particle pairs (i.e.,\ an $\epsilon$ neighborhood of the points $(\pm 1, 0)$), this configuration is where $g(x,y)$ is sharply peaked, as was shown in Fig.\ \ref{structure}. We show $\fmov$ for $5\times 10^4$ isolated pairs of particles that initially overlap, trained with cyclic shear of strain amplitude $\gamma_1=2$ in Fig.\ \ref{memorystep}(b). The size of the jump is less than unity because some pairs separate such that they will never collide, even for arbitrarily large shear amplitudes (by separating either vertically, or into the first or third quadrants of the domain of $g(x,y)$) .

Finally, three-body effects smooth the step function even more. In Fig.\ \ref{memorystep}(c), we show $\fmov$ for a single system with $N=10^5$ particles and area fraction $\phi=0.2$ driven at $\gamma_1=2$ to a reversible state, under the Cort\'e algorithm. (We used $N=10^4$ particles for $\epsilon=0.001$, due to the prohibitively long computing times of a larger system.) As $\epsilon$ decreases, $\fmov$ becomes steeper at $\gamma_1$, but the curve does not have a punctuated jump, except perhaps at the smallest kick size tested, $\epsilon=0.001$.

We have given a qualitative account of the disappearance of the memory step function in $\fmov$ for a complete memory. To be more quantitative, we articulate two measures of the degree to which a particular $\fmov$ curve is approximated by a step function. The first measure is the size of the jump in $\fmov$ for a test shear to $\gamma_1$ versus a test shear to $\gamma_1+2\epsilon$. For a complete memory, the size of the jump is simply $\fmov(\gamma_1+2\epsilon)$, since $\fmov(\gamma_1)=0$. We show this value versus $\epsilon$ in Fig.\ \ref{memorystep}(d). Whereas swelling of isolated pairs of particles gives a perfect step (i.e.,\ a jump of unity for all values of $\epsilon$), shearing of isolated pairs and including three-body interactions limit the jump to a smaller value for small $\epsilon$. 

The second measure is the slope of $\fmov$ on the right-hand-side of $\gamma_1$, denoted by $\fmov'(\gamma_1)_+$. 
For small enough $\epsilon$, the dominant length scale that dictates this slope will be $\epsilon$, due to a separation in length scales between the kick size, $\epsilon$, and the particle size. We plot the product $\epsilon \fmov'(\gamma_1)_+$ in Fig.\ \ref{memorystep}(e). The data plateau to a finite value for small $\epsilon$ under all three algorithms, indicating that the slope approaches this prefactor times $1/\epsilon$. The prefactor indicates how well $\fmov$ is approximated by a step function, which increases as we take out three-body effects and change to a swelling system.

Both quantitative measures show that there is a step-function contribution to  $\fmov$ in all cases.  However, this step function is broadened by at least two effects. Our results indicate
that three-body effects are the smaller contribution to the broadening and shortening of the step function, when compared with the effect of the cusp in the excluded region for the shearing algorithm. This suggests that an experimental system that employs particle swelling may have a sharper signature of memory formation (assuming all other effects stay constant).

\subsection{Faster training with $\gamma=3,2,2,2,2,2$ is an artifact of drift}

Finally, we note one surprising feature found when training two memories in the sheared suspension simulations. In ref.~\cite{Keim2011}, when a system of $N=10^4$ particles with area fraction $\phi=0.2$ was trained simultaneously at two strain amplitudes, $\gamma_1=2$ and $\gamma_2=3$, with the pattern of shears, $3,2,2,2,2,2$, the system organized to a fully reversible state in $\sim 30\,000$ cycles, compared with $\sim 55\,000$ cycles if the system was trained with just a single amplitude, $\gamma_1=3$. We have investigated this effect further, and we find that the effect is statistically significant, and grows with system size (we tested system sizes from $N=11$ to $N=10^5$). The effect is  more pronounced in Variant~\ref{variant:tko} (``Tag-kick-once"), versus Variant~\ref{variant:corte} (original kinematics of Cort\'e \textit{et al.}). 

We find that the effect may be attributed to the ability of particle pairs to drift: since in those two algorithms two overlapping particles each receive a random kick in a random direction, their center of mass may diffuse. This mechanism speeds up training under a more ``gentle'' protocol of $\gamma_1=2$ interspersed with $\gamma_2=3$, which allows higher-density regions to relax while minimizing the effect of particle pairs wandering off and disturbing regions that had already been trained to a reversible state. When we incorporate momentum conservation into these algorithms (Variant~\ref{variant:momcons}, and another variant similar to ``Tag-kick-once'' and including momentum conservation) so that diffusion of interacting particles is suppressed, we find that the effect is almost completely eliminated.


\section{Driving above the threshold for irreversibility} 
\label{theme3}

\begin{figure}[bt] 
\includegraphics[width=3.4in]{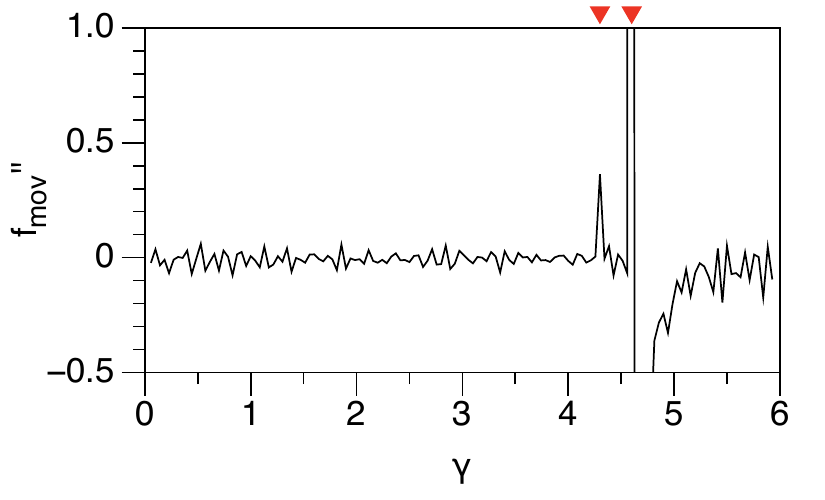}
\caption{
(color online) Multiple-memory behavior at strain values above the critical threshold for irreversibility. Here $\gamma_c = 4.0$; $\gamma_1 = 4.3$ and $\gamma_2 = 4.6$ (red triangles at top). Driving above $\gamma_c$ prevents the system from reaching a reversible steady state with only one memory, and so both training inputs are retained indefinitely. The smaller shear value decays negligibly after $\sim 2 \times 10^6$ cycles. The curve plotted here is from cycle $5.0 \times 10^6$ and is an average over 100 initial conditions.
\label{fig:over_memories}
}
\end{figure}

While we have treated different training values of $\gamma$ as more or less equivalent, in fact as $\gamma$ is increased, it takes longer and longer for the system to reach a reversible steady state. Cort\'e \textit{et al.}~\cite{Corte2008} demonstrated that the characteristic time for self-organization diverges at a value $\gamma_c$; this is a critical transition in the conserved directed percolation class~\cite{Menon2009}. For $\gamma_1 > \gamma_c$, reversible (non-interacting) arrangements of the particles may exist, but the system cannot reach them through its self-organization process~\cite{Corte2008}. In the simulations of memory effects reported above and in ref.~\cite{Keim2011}, all $\{\gamma_i\}$ were kept below $\gamma_c$, so that in the absence of external noise, the system would always evolve to a reversible steady state with only a single memory.

However, if we permit one or more $\gamma_i > \gamma_c$ in the training protocol, the system can never reach a reversible steady state; some particles are kicked in each cycle. This restriction bears a resemblance to that imposed by external noise~\cite{Povinelli1999}. Keim and Nagel~\cite{Keim2011} showed that when uncorrelated noise is added to the particles' positions on each cycle, with a scale $\epsilon_\text{noise} \ll \epsilon$, self-organization to reversibility is averted and multiple memories may persist indefinitely. Similarly, we expect that in the absence of noise, a system driven with some $\gamma_i > \gamma_c$ should sustain multiple memories indefinitely, if it supports the formation of multiple memories at all. Because our simulations must run for finite time, we consider memories to be phenomenologically indefinite when they do not decay over timescales much longer than the time required to enter an apparent steady state; in this regime, changes in memory strength are dominated by fluctuations and no decay is discernible. 

\subsection{Forming Memories Above $\gamma_c$} 

Figure~\ref{fig:over_memories} shows that multiple memories are indeed possible with $\gamma_i > \gamma_c$. We use the same parameters and kinematics as in Fig.~\ref{fig:variants}(b) (Variant~\ref{variant:tko}). The training strain amplitudes are $\gamma_1 = 4.3, \gamma_2 = 4.6$, with training pattern $\gamma_1, \gamma_2, \gamma_2, \gamma_2, \gamma_2, \gamma_2$, repeat\dots, and $\gamma_c = 4.0$. We observe that after $\sim 2 \times 10^6$ cycles, both memories appear to be stable and present in each cycle, in contrast to the ``forgetting'' behavior reported for $\{\gamma_i\} < \gamma_c$ in ref.~\cite{Keim2011}.  (To compute the curve, $\fmov$ was sampled at intervals of $\Delta \gamma = 0.0429$, so that sampling positions coincided with both training values.)

\subsection{Overdriving as Noise} 

\begin{figure}[bt] 
\includegraphics[width=3.4in]{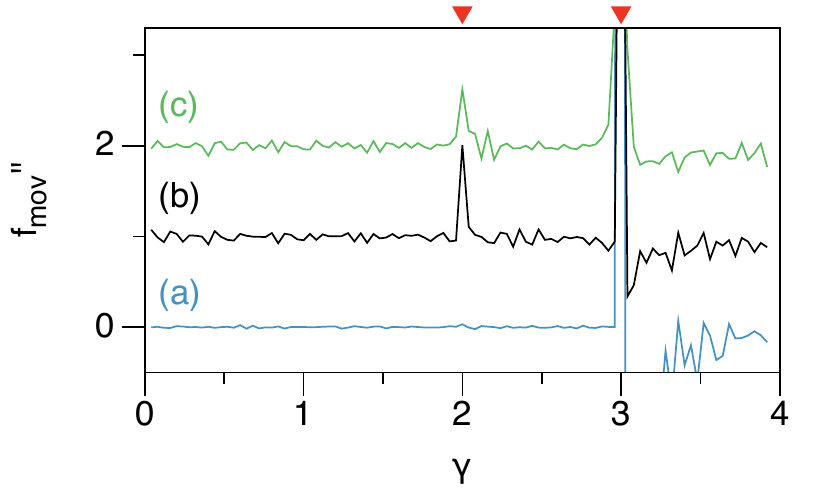}
\caption{
(color online) Multiple memories retained by occasional application of a large strain. Training pattern is $\gamma_i =$ 3, 2, 2, 2, 2, 2 (red triangles at top). Parameters and kinematics are the same as in Fig.~\ref{fig:variants}(b), but with $\epsilon = 0.015$. Curves are offset by $1$ for clarity.
    \textbf{(a)} Without application of any strain greater than $\gamma_c = 4.0$, the $\gamma_1 = 2$ memory is essentially lost after $7 \times 10^5$ cycles shown here (average of 48 runs). It is completely gone (by definition) when the simulation reaches a steady state that is reversible for $\gamma_1 = 3$ ($\sim1.2 \times 10^6$ cycles). 
    \textbf{(b, c)} Simulation with $\gamma_\text{large} = 100$ applied every $\sim 100$ cycles. Large-amplitude shearing takes the role of noise, preventing organization to a reversible steady state and so indefinitely maintaining plasticity, the system's susceptibility to multiple memories. Curves show memories after $\sim 2 \times 10^6$ cycles, (b) before and (c) after application of the large-amplitude strain. The peaks in (c) at the two training amplitudes are still well defined showing that this protocol ``refreshes'' the system without destroying memories.
Each curve is an average over 96 runs.
\label{fig:over_noise}
}
\end{figure}

A second, more practical use of $\gamma > \gamma_c$ is to simply take the role of external noise in a training protocol that otherwise has only $\{\gamma_i\} < \gamma_c$. We have realized this principle by intermittently applying a $\gamma_\text{large} \gg \gamma_c$, which for suitably small $\epsilon$ and infrequent application, disrupts the particles' self-organized positions to prevent a reversible steady state, but not so much as to completely destroy already-formed memories. We wish to apply this large shear every $\Delta t_\text{large}$ cycles, to mimic noise applied every cycle with a scale $\sim 0.1 \epsilon$, which is known to sustain memories indefinitely~\cite{Keim2011}. We use the scaling of displacement for a random walk, $\Delta x \sim t^{1/2}$, to arrive at $\Delta t_\text{large} \sim 100$. Since in Variant~\ref{variant:tko} of the kinematics multiple collisions in a cycle do not add to displacements, our large strain amplitude need only satisfy $\gamma_\text{large} \gg \gamma_c$; here we choose $\gamma_\text{large} = 100$. For kinematics in which multiple collisions are additive, such as Variant~\ref{variant:corte}, $\gamma_\text{large}$ would have to be selected more carefully to avoid erasure of existing memories, using the considerations outlined in sec.~\ref{sec:gradual}.

The results of overdriving used as noise are shown in Fig.~\ref{fig:over_noise}, where multiple memories are sustained indefinitely, long after an equivalent simulation without large-amplitude shearing reaches a steady state with just one memory. This technique has the advantage that the sustaining ``noise'' is added in the same way that the driving signal is applied, rather than relying on an additional noise mechanism that may be difficult to introduce or control.


\section{Conclusion} 

In summary, we have explored multiple transient memories in simulations of sheared suspensions, focussing on three main themes: robustness, structure, and overdriving. We have shown that multiple transient memories are a robust feature that are manifest in a variety of very simple models of suspensions under cyclic, low Reynolds-number shear.  The details of memory formation can be understood from the spatial correlations of the particles.  

We find that overdriving can provide another means for controlling memory formation and retention.  Remarkably, memories can be stored, not only for training at $\gamma<\gamma_c$ but also at amplitudes $\gamma>\gamma_c$.  Moreover, we have found that a large shear that exceeds the critical value, $\gamma_c$, can be used as an alternative to noise to allow memories to persist indefinitely. 

As demonstrated by Keim \& Nagel \cite{Keim2011} and recapitulated over a wider range of algorithm variants in this work, the requirements for transient memories are relatively minimal. The salient components appear to be a system that relaxes to a reversible state under cyclic driving during a transient period of irreversibility, and an ordering of reversible states, meaning that a system reversible under a strain amplitude $\gamma_1$ is reversible under smaller shears, $\gamma<\gamma_1$. Thus, we expect other sheared disordered systems with these properties to exhibit transient memories as well, such as granular materials \cite{Toiya2004, Mueggenburg2005}, colloids \cite{Ackerson1988, Haw1998}, foams \cite{Lundberg2008}, and filament networks \cite{schmoller10}, and under other driving protocols, such as tapping \cite{Knight1995}. These results, combined with the previous studies of charge-density waves, suggest that multiple transient memories may be a generic feature in a wide class of driven disordered systems.

\begin{acknowledgments} 
We thank Thomas A.\ Caswell for assistance in optimizing the simulations. This work was supported by NSF Grant DMR-1105145 and by NSF MRSEC DMR-0820054 (J.D.P.\ and S.R.N.), and by NSF MRSEC DMR-1120901 (N.C.K.).  Use of computation facilities funded by the US Department of Energy, Office of Basic Energy Sciences, Division of Materials Sciences and Engineering, Award No.\ DE-FG02-03ER46088 is gratefully acknowledged.
\end{acknowledgments}

\bibliography{MemoriesBib}

\end{document}